\documentclass[aps,pra,reprint]{revtex4-2}
\usepackage{graphicx, amsmath, amssymb, hyperref}
\usepackage[caption=false]{subfig}

\begin{document}

%-------------------------------------------------------------------------------
% Shortcut Commands
%-------------------------------------------------------------------------------

\newcommand{\comb}[2]{{\begin{pmatrix} #1 \\ #2 \end{pmatrix}}}
\newcommand{\braket}[2]{{\left\langle #1 \middle| #2 \right\rangle}}
\newcommand{\bra}[1]{{\left\langle #1 \right|}}
\newcommand{\ket}[1]{{\left| #1 \right\rangle}}
\newcommand{\ketbra}[2]{{\left| #1 \middle\rangle \middle \langle #2 \right|}}

\newcommand{\fref}[1]{Fig.~\ref{#1}}
\newcommand{\Fref}[1]{Figure~\ref{#1}}
\newcommand{\sref}[1]{Sec.~\ref{#1}}
\newcommand{\tref}[1]{Table.~\ref{#1}}

%-------------------------------------------------------------------------------
% Front Matter
%-------------------------------------------------------------------------------

\title{Searching Weighted Barbell Graphs with Laplacian and Adjacency Quantum Walks}

\author{Jonas Duda}
	\email{jonasduda@creighton.edu}
	\affiliation{Department of Physics, Creighton University, 2500 California Plaza, Omaha, NE 68178}
\author{Thomas G. Wong}
	\email{thomaswong@creighton.edu}
	\affiliation{Department of Physics, Creighton University, 2500 California Plaza, Omaha, NE 68178}

\begin{abstract}
	A quantum particle evolving by Schr\"odinger's equation in discrete space constitutes a continuous-time quantum walk on a graph of vertices and edges. When a vertex is marked by an oracle, the quantum walk effects a quantum search algorithm. Previous investigations of this quantum search algorithm on graphs with cliques have shown that the edges between the cliques can be weighted to enhance the movement of probability between the cliques to reach the marked vertex. In this paper, we explore the most restrictive form of this by analyzing search on a weighted barbell graph that consists of two cliques of the same size joined by a single weighted edge/bridge. This graph is generally irregular, so quantum walks governed by the graph Laplacian or by the adjacency matrix can differ. We show that the Laplacian quantum walk's behavior does not change, no matter the weight of the bridge, and so the single bridge is too restrictive to affect the walk. Similarly, the adjacency quantum walk's behavior is unchanged for most weights, but when the weight equals the size of a clique, the probability does collect at the clique containing the marked vertex, and utilizing a two-stage algorithm with different weights for each stage, the success probability is boosted from 0.5 to 0.996, independent of the size of the barbell graph.
\end{abstract}

\maketitle

%-------------------------------------------------------------------------------
% Main Matter
%-------------------------------------------------------------------------------

\section{Introduction}

In quantum mechanics \cite{Griffiths2018}, a particle with wave function $\psi$ evolves by Schr\"odinger's equation:
\begin{equation}
    \label{eq:Schrodinger}
    i \frac{\partial \psi}{\partial t} = H \psi,
\end{equation}
where we have set $\hbar = 1$. The Hamiltonian $H$ is the operator corresponding to the total energy of the system, and in general,
\[ H = -\frac{1}{2m} \nabla^2 + V, \]
where in 3-dimensional Cartesian coordinates, $\nabla^2 = \partial^2/\partial x^2 + \partial^2/\partial y^2 + \partial/\partial z^2$ is the Laplace operator, $-\nabla^2/2m$ is the kinetic energy of the particle, and $V$ is the potential energy.

In discrete space, which can be modeled as an undirected graph of $N$ vertices and edges, the particle can only be at the vertices, and it can only jump along edges. This is known as a continuous-time quantum walk, and is the quantum analogue of a continuous-time random walk \cite{Wong35}. The vertices correspond to computational basis states $\ket{1}, \ket{2}, \dots, \ket{N}$, and the state $\ket{\psi}$ of the particle is generally a superposition over the vertices. In the Hamiltonian, the continuous-space Laplace operator $\nabla^2$ is replaced by the discrete Laplacian
\begin{equation}
    \label{eq:L}
    L = A - D,
\end{equation}
where $A$ is the adjacency matrix of the graph ($A_{ij} = 1$ if vertices $i$ and $j$ are adjacent, and 0 otherwise), and $D$ is the diagonal degree matrix with $D_{ii}$ equal to the degree of vertex $i$, i.e., its number of neighbors. Using this and writing $\gamma = 1/2m$, the Hamiltonian becomes
\begin{equation}
    \label{eq:H-Laplacian}
    H_L = -\gamma L + V.
\end{equation}
We refer to this as a \emph{Laplacian quantum walk}. Laplacian quantum walks were introduced for traversing decision trees \cite{FG1998b}, and they have also been used for state transfer \cite{Alvir2016} and spatial search \cite{CG2004}.

In some cases, such as a system of interacting spins with $XY$ interactions and single excitations \cite{Bose2009}, the quantum walk is effected by the adjacency matrix of the graph rather than the full discrete Laplacian. Then, the Hamiltonian is
\begin{equation}
    \label{eq:H-Adjacency}
    H_A = -\gamma A + V.
\end{equation}
We refer to this as an \emph{adjacency quantum walk}. Adjacency quantum walks provide an exponential speedup in traversing glued trees \cite{Childs2003}, and they has also been used for solving boolean formulas \cite{FGG2008}, state transfer \cite{Godsil2012}, and spatial search \cite{Novo2015}.

In general, Laplacian and adjacency quantum walks evolve differently \cite{Wong37} and search differently \cite{Wong19}. One situation where they are equivalent, however, is when the graph is regular, meaning each vertex has the same degree. Then, the diagonal degree matrix $D$ is proportional to the identity matrix, and it can be dropped by rezeroing the energy. This manifests itself as a physically irrelevant global phase that can be dropped, i.e., the evolution of the Laplacian and adjacency quantum walks are equal up to a global phase. Some weighted graphs---where the terms in the adjacency matrix $A_{ij}$ take the weight of the edge between vertices $i$ and $j$, and the degree of each vertex is the sum of the weights of its edges---are regular, and the quantum walks are equivalent on these regular weighted graphs as well.

\begin{figure}
\begin{center}
    \includegraphics{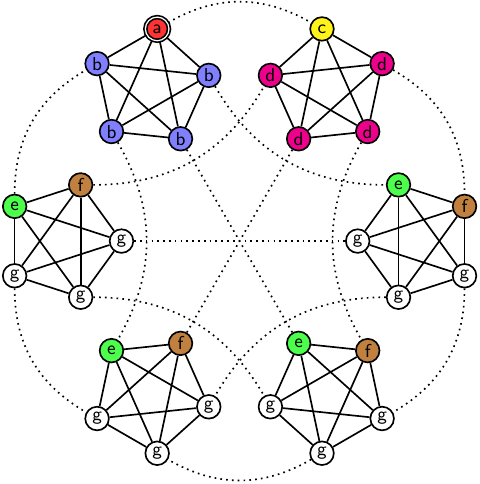}
    \caption{\label{fig:truncated}A weighted first-order truncated simplex lattice of $N = 30$ vertices, which contains 6 cliques with $M = 5$ vertices each. Solid edges are unweighted, and dotted edges have weight $w$. A vertex is marked, indicated by a double circle. Identically evolving vertices are identically colored and labeled.}
\end{center}
\end{figure}

For example, the two quantum walks are equivalent when searching a weighted (first-order) \emph{truncated simplex lattice} \cite{Wong16}, an example of which is shown in \fref{fig:truncated}, for a vertex marked by an oracle, i.e., a \emph{marked vertex}. It consists of $M+1$ cliques, each of size $M$, such that each vertex in a clique is adjacent to one other unique clique through a weighted edge. In \fref{fig:truncated}, the solid edges within the cliques have weight $1$, and the dotted edges between the cliques have variable weight $w$. The total number of vertices is $N = M(M+1)$. The degree of each vertex is $M + w - 1$, so it is a regular graph, and the Laplacian and adjacency quantum walks are equivalent. As analytically shown in \cite{Wong16}, as the weight $w$ increases, search on the truncated simplex lattice speeds up, going from a runtime of $O(N^{3/4})$ to $o(\sqrt{N})$, i.e., nearly a scaling of $\sqrt{N}$. In \cite{Wang2020}, it was numerically shown that $w$ can be increased further to achieve a runtime of $\Theta(\sqrt{N})$, i.e., a scaling of $\sqrt{N}$.

\begin{figure}
\begin{center}
    \includegraphics{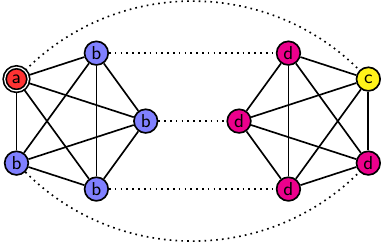}
    \caption{\label{fig:linked}Weighted linked complete graphs with $N = 10$ vertices. Solid edges are unweighted, and dotted edges have weight $w$. A vertex is marked, indicated by a double circle. Identically evolving vertices are identically colored and labeled.}
\end{center}
\end{figure}

The quantum walks are also equivalent when searching \emph{linked complete graphs}, which consists of two cliques of equal size such that each vertex in one clique is adjacent to a unique vertex in the other clique, an example of which is shown in \fref{fig:linked}. Formally, this is the Cartesian product of the complete graphs $K_2$ and $K_{N/2}$. As before, the solid edges within the cliques are unweighted, and the dotted edges between the cliques have weight $w$. Then, the degree of each vertex is $N/2 + w - 1$, so it is a regular graph, and the Laplacian and adjacency quantum walks are equivalent. As shown in \cite{Wong22}, when $w = 1$, the success probability (i.e., the probability that measuring the particle's position yields the marked vertex) reaches $1/2$ in time $\pi\sqrt{N}/2\sqrt{2}$. To explain this, let us refer to the clique that contains the marked vertex (i.e., the left clique in \fref{fig:linked}) as the \textit{marked clique}, and the other clique (i.e., the right clique in \fref{fig:linked}) as the \textit{unmarked clique}. The quantum search algorithm begins in a uniform superposition over all $N$ vertices, so each clique begins with half the probability, and as the algorithm evolves, the probability in the marked clique gathers at the marked vertex, while the probability in the unmarked clique is trapped. That is, the algorithm behaves like search on a single complete graph \cite{CG2004,Wong35} of $N/2$ vertices with half the probability. As shown in \cite{Wong22}, as $w$ increases, however, the trapped probability is able to escape, and when $w$ scales greater than $\sqrt{N}$ but less than $N$, it is fully able to escape, and the success probability reaches 1 for large $N$. If $w$ is increased further, however, the success probability gets worse. Thus, there is a region of $w$ where the algorithm is optimized.

\begin{figure}
\begin{center}
    \includegraphics{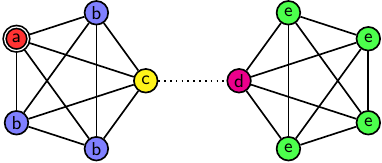}
    \caption{\label{fig:barbell}A weighted barbell graph of $N = 10$ vertices. Solid edges are unweighted, and the dotted edge has weight $w$. A vertex is marked, indicated by a double circle. Identically evolving vertices are identically colored and labeled.}
\end{center}
\end{figure}

In these two examples, the weights of the edges between the cliques could be tuned in order to speed up the algorithm or increase the success probability. These raise questions on how weighted edges between cliques can be exploited for search on other graphs. In this paper, we consider a weighted \emph{barbell graph}, an example of which is shown in \fref{fig:barbell}. It consists of two cliques of size $N/2$, joined by a single weighted edge or bridge, so it has the fewest number of edges possible while keeping the two cliques connected. The unweighted ($w = 1$) version of this graph was analyzed in \cite{Wong7}, and it was shown that the probability in the unmarked clique is trapped, just like in the linked complete graphs, so the algorithm behaves like a single complete graph with half the vertices and probability. In this paper, we investigate whether increasing the weight $w$ can free the probability in the unmarked clique, or if having a single edge is too restrictive to help.

Besides being more restrictive, the barbell graph differs in an important way from the truncated simplex lattice and the linked complete graphs: it is irregular. In \fref{fig:barbell}, vertices $c$ and $d$ have degree $N/2 - 1 + w$, whereas the other vertices have degree $N/2 - 1$. For large $N$, if $w$ scales less than $N$ [i.e., $w = o(N)$], then $N/2 - 1 + w$ and $N/2 - 1$ are both approximately $N/2$, meaning the graph is approximately regular in this case, and the Laplacian and adjacency quantum walks are asymptotically equivalent. If $w$ scales at least as large as $N$ [i.e., $w = \Omega(N)$], then $N/2 - 1 + w$ and $N/2 - 1$ are not asymptotically equivalent, and the two quantum walks will differ.

In the next section, we detail the quantum walk search algorithm on the weighted barbell graph and show that the system evolves in a 5-dimensional subspace. We give the initial state, adjacency matrix, degree matrix, Laplacian, and potential energy in this subspace, as well as the Hamiltonians for the Laplacian and adjacency quantum walks. We numerically explore how each quantum walk behaves to motivate the analytical proofs in the subsequent sections. In \sref{sec:Laplacian}, we analytically prove how the Laplacian quantum walk searches the weighted barbell graph, and we show that no matter the value of $w$, the system asymptotically behaves like search on the unweighted graph, reaching a success probability of $1/2$. Then, there is no value of $w$ that can free the trapped probability in the unmarked clique, and having a single bridge between the cliques is too restrictive to help. In \sref{sec:adjacency}, we analyze the adjacency quantum walk, and we show that for almost all values of $w$, this also behaves like search on the unweighted graph. The exception to this is when $w = N/2$, at which probability is able to escape from the unmarked clique and collect at the marked vertex, reaching a success probability of 0.820. In \sref{sec:twostage}, we analyze a two-stage algorithm that uses the adjacency quantum walk's behavior when $w = N/2$ for the first stage, followed by the unweighted graph's behavior for the second stage, to achieve a success probability of 0.996, which is near certainty, regardless of the size of the barbell graph. In \sref{sec:conclusion}, we conclude.

%-------------------------------------------------------------------------------
% Section
%-------------------------------------------------------------------------------

\section{\label{sec:algorithm}Search Algorithm}

In quantum walk search algorithms \cite{CG2004}, the system begins in a uniform superposition over the vertices, which in the $N$-dimensional computational basis is
\[ \ket{\psi(0)} = \frac{1}{\sqrt{N}} \sum_{i=1}^N \ket{i}. \]
In the search algorithm, the potential energy is
\begin{equation}
    \label{eq:V}
    V = -\ketbra{a}{a},
\end{equation}
where $\ket{a}$ is the marked vertex that we are searching for, and this potential energy acts as the oracle \cite{CG2004}. The solution to Schr\"odinger's equation gives the state of the system at time $t$, and since the Hamiltonians for the Laplacian \eqref{eq:H-Laplacian} and adjacency \eqref{eq:H-Adjacency} quantum walks are time-independent, the solution is
\begin{equation}
    \label{eq:psi}
    \ket{\psi(t)} = e^{-iHt} \ket{\psi(0)}.
\end{equation}
The amplitude at the marked vertex at time $t$ is $\braket{a}{\psi(t)}$, and taking the norm-square of this, the success probability is
\begin{equation}
    \label{eq:prob-general}
    p_a(t) = \left| \braket{a}{\psi(t)} \right|^2.
\end{equation}

As in \cite{Wong7}, we can reduce the dimension of the space using the symmetry of the barbell graph. As labeled in \fref{fig:barbell}, the barbell graph has five types of vertices, $a$, $b$, $c$, $d$, and $e$, where $a$ is the marked vertex, $c$ and $d$ are the endpoints of the bridge ($c$ in the marked clique and $d$ in the unmarked one), $b$ are the vertices in the marked clique except for $a$ and $c$, and $e$ are the vertices in the unmarked clique, except for $d$. Each vertex of the same type evolves identically, e.g., as the system evolves, the amplitude at each of the $b$ vertices is the same. So, the system evolves in a 5D subspace spanned by the orthonormal basis consisting of uniform superpositions over each type of vertex:
\begin{align*}
	&\ket{a}, \vphantom{\frac{1}{\sqrt{N}}} \displaybreak[0] \\
	&\ket{b} = \frac{1}{\sqrt{N/2 - 2}} \sum_{i \in b} \ket{i}, \displaybreak[0] \\
	&\ket{c}, \vphantom{\frac{1}{\sqrt{N}}} \displaybreak[0] \\
	&\ket{d}, \vphantom{\frac{1}{\sqrt{N}}} \displaybreak[0] \\
	&\ket{e} = \frac{1}{\sqrt{N/2 - 1}} \sum_{i \in e} \ket{i}.
\end{align*}
In this $\{ \ket{a}, \ket{b}, \dots, \ket{e} \}$ basis, the initial uniform state is
\begin{align}
    \ket{\psi(0)} 
        &= \frac{1}{\sqrt{N}} \ket{a} + \frac{\sqrt{N/2 - 2}}{\sqrt{N}} \ket{b} + \frac{1}{\sqrt{N}} \ket{c} \nonumber \\
        &\quad+ \frac{1}{\sqrt{N}} \ket{d} + \frac{\sqrt{N/2 - 1}}{\sqrt{N}} \ket{e}, \label{eq:psi0}
\end{align}
or as a column vector,
\begin{equation}
    \label{eq:psi0-colvec}
    \ket{\psi(0)} = \frac{1}{\sqrt{N}} \begin{pmatrix}
        1 \\
        \sqrt{\frac{N}{2} - 2} \\
        1 \\
        1 \\
        \sqrt{\frac{N}{2} - 1} \\
    \end{pmatrix}.
\end{equation}
Similarly, the adjacency matrix is
\begin{widetext}
\[ A = \begin{pmatrix}
	0 & \sqrt{\frac{N}{2}-2} & 1 & 0 & 0 \\
	\sqrt{\frac{N}{2}-2} & \frac{N}{2} - 3 & \sqrt{\frac{N}{2}-2} & 0 & 0 \\
	1 & \sqrt{\frac{N}{2}-2} & 0 & w & 0 \\
	0 & 0 & w & 0 & \sqrt{\frac{N}{2}-1} \\
	0 & 0 & 0 & \sqrt{\frac{N}{2}-1} & \frac{N}{2}-2 \\
\end{pmatrix}, \]
where, for example, $A_{12}$ comes from the $1$ vertex of type $a$ that is adjacent to a type $b$ vertex, multiplied by $\sqrt{N/2-2}/\sqrt{1}$ to convert between the normalizations of the $\ket{b}$ and $\ket{a}$ basis states. Similarly, $A_{21}$ comes from the $N/2-2$ vertices of type $b$ that are adjacent to the type $a$ vertex, times $\sqrt{1}/\sqrt{N/2-2}$ to convert between the normalizations of the $\ket{a}$ and $\ket{b}$ basis states. Next, the degree matrix is
\[ D = \begin{pmatrix}
    \frac{N}{2} - 1 & 0 & 0 & 0 & 0 \\
    0 & \frac{N}{2} - 1 & 0 & 0 & 0 \\
    0 & 0 & \frac{N}{2} - 1 + w & 0 & 0 \\
    0 & 0 & 0 & \frac{N}{2} - 1 + w & 0 \\
    0 & 0 & 0 & 0 & \frac{N}{2} - 1 \\
\end{pmatrix}, \]
so the Laplacian $L = A - D$ \eqref{eq:L} is
\[ L = \begin{pmatrix}
	-\frac{N}{2} + 1 & \sqrt{\frac{N}{2}-2} & 1 & 0 & 0 \\
	\sqrt{\frac{N}{2}-2} & -2 & \sqrt{\frac{N}{2}-2} & 0 & 0 \\
	1 & \sqrt{\frac{N}{2}-2} & -\frac{N}{2}+1-w & w & 0 \\
	0 & 0 & w & -\frac{N}{2} + 1 - w & \sqrt{\frac{N}{2}-1} \\
	0 & 0 & 0 & \sqrt{\frac{N}{2}-1} & -1 \\
\end{pmatrix}. \]
Finally, the potential energy \eqref{eq:V}, or oracle, is
\[ V = \begin{pmatrix}
    -1 & 0 & 0 & 0 & 0 \\
    0 & 0 & 0 & 0 & 0 \\
    0 & 0 & 0 & 0 & 0 \\
    0 & 0 & 0 & 0 & 0 \\
    0 & 0 & 0 & 0 & 0 \\
\end{pmatrix}. \]
Putting these together using \eqref{eq:H-Laplacian}, the Hamiltonian for the Laplacian quantum walk search algorithm is
\begin{equation}
    \label{eq:H-Laplacian-5D}
    H_L = -\gamma \begin{pmatrix}
        -\frac{N}{2} + 1 + \frac{1}{\gamma} & \sqrt{\frac{N}{2}-2} & 1 & 0 & 0 \\
        \sqrt{\frac{N}{2}-2} & -2 & \sqrt{\frac{N}{2}-2} & 0 & 0 \\
        1 & \sqrt{\frac{N}{2}-2} & -\frac{N}{2}+1-w & w & 0 \\
        0 & 0 & w & -\frac{N}{2} + 1 - w & \sqrt{\frac{N}{2}-1} \\
        0 & 0 & 0 & \sqrt{\frac{N}{2}-1} & -1 \\
    \end{pmatrix},
\end{equation}
and using \eqref{eq:H-Adjacency}, the Hamiltonian for the adjacency quantum walk search algorithm is
\begin{equation}
    \label{eq:H-Adjacency-5D}
    H_A = -\gamma \begin{pmatrix}
        \frac{1}{\gamma} & \sqrt{\frac{N}{2}-2} & 1 & 0 & 0 \\
        \sqrt{\frac{N}{2}-2} & \frac{N}{2} - 3 & \sqrt{\frac{N}{2}-2} & 0 & 0 \\
	    1 & \sqrt{\frac{N}{2}-2} & 0 & w & 0 \\
        0 & 0 & w & 0 & \sqrt{\frac{N}{2}-1} \\
        0 & 0 & 0 & \sqrt{\frac{N}{2}-1} & \frac{N}{2}-2 \\
    \end{pmatrix}.
\end{equation}
\end{widetext}

\begin{figure}
\begin{center}
    \includegraphics{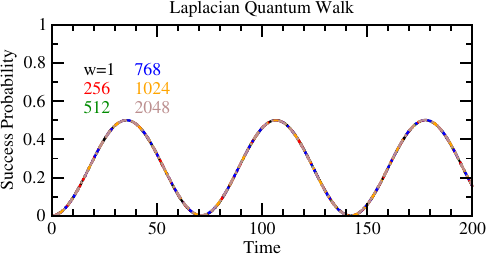}
    \caption{\label{fig:Laplacian-N1024}Success probability versus time for search on the barbell of graph of $N = 1024$ vertices using a Laplacian quantum walk with $\gamma = 2/N$. The solid black curve is $w = 1$, the dashed red curve is $w = 256$, the dotted green curve is $w = 512$, the dot-dashed blue curve is $w = 768$, the dot-dot-dashed orange curve is $w = 1024$, and the dot-dashed-dashed brown curve is $w = 2048$. All the curves are on top of each other.}
\end{center}
\end{figure}
Using these Hamiltonians and the initial state \eqref{eq:psi0-colvec}, we can find the state at time $t$ using \eqref{eq:psi}, then take the inner product with $\bra{a}$ to find the success amplitude, and then take the norm square to get the success probability \eqref{eq:prob-general}. We numerically plot this success probability as a function of time in \fref{fig:Laplacian-N1024} for the Laplacian quantum walk on the weighted barbell graph with $N = 1024$ vertices and six different values of the weight: $w = 1$, $256$, $512$, $768$, $1024$, and $2048$. We see that no matter the weight, the behavior is the same as the unweighted version, i.e., the success probability reaches $1/2$ at time $\pi\sqrt{N}/2\sqrt{2} \approx 35.5$. This is because the probability in the marked clique collects at the marked vertex, while the probability in the unmarked clique is trapped. 

\begin{figure}
\begin{center}
    \includegraphics{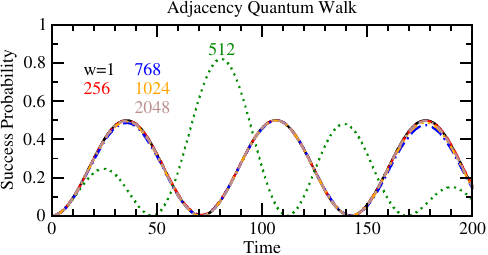}
    \caption{\label{fig:Adjacency-N1024}Success probability versus time for search on the barbell of graph of $N = 1024$ vertices using an adjacency quantum walk with $\gamma = 2/N$. The solid black curve is $w = 1$, the dashed red curve is $w = 256$, the dotted green curve is $w = 512$, the dot-dashed blue curve is $w = 768$, the dot-dot-dashed orange curve is $w = 1024$, and the dot-dashed-dashed brown curve is $w = 2048$. The $w = 512$ curve stands above the others, while the rest are on top of each other.}
\end{center}
\end{figure}

\begin{figure}
\begin{center}
    \includegraphics{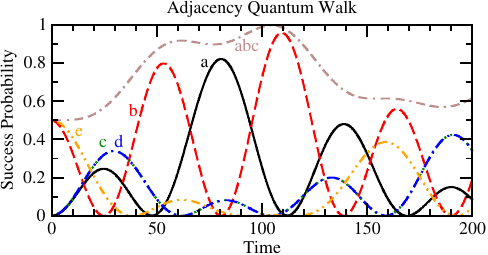}
    \caption{\label{fig:Adjacency-N1024-w512-types}Search on the barbell of graph of $N = 1024$ vertices using an adjacency quantum walk with $\gamma = 2/N$ and $w = 512$. The solid black curve is the probability of finding the particle at the marked $a$ vertex (which is the same as the dotted green curve in \fref{fig:Adjacency-N1024}), the dashed red curve is a $b$ vertex, the dotted green curve is the $c$ vertex, the dot-dashed blue curve is the $d$ vertex, the dot-dot-dashed orange curve is an $e$ vertex, and the dot-dashed-dashed brown curve is the sum of the probabilities for the $a$, $b$, and $c$ vertices, i.e., the marked clique.}
\end{center}
\end{figure}

\begin{figure}
\begin{center}
    \includegraphics{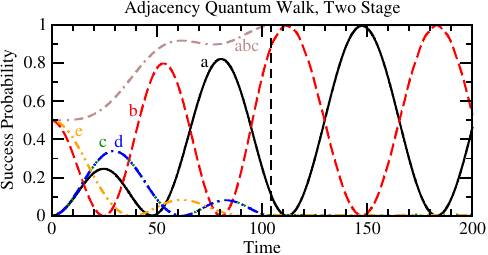}
    \caption{\label{fig:Adjacency-twostage-N1024}Search on the barbell of graph of $N = 1024$ vertices using an adjacency quantum walk with $\gamma = 2/N$. When $t < 104.4$, as indicated by the black dashed vertical line, $w = 512$, and when $t \ge 104.4$, $w = 1$. The solid black curve is the probability of finding the particle at the marked $a$ vertex (which is the same as the dotted green curve in \fref{fig:Adjacency-N1024}), the dashed red curve is a $b$ vertex, the dotted green curve is the $c$ vertex, the dot-dashed blue curve is the $d$ vertex, the dot-dot-dashed orange curve is an $e$ vertex, and the dot-dashed-dashed brown curve is the sum of the probabilities for the $a$, $b$, and $c$ vertices, i.e., the marked clique.}
\end{center}
\end{figure}

Similarly, for the adjacency quantum walk, we get \fref{fig:Adjacency-N1024}. We see that, for most values of $w$, the search behaves just like the unweighted case. When $w = 512$, however, the behavior is different, and the success probability is boosted to roughly 0.820 at time 80.6. From this, one might conclude that a single weighted bridge is \mbox{insufficient} to fully free the probability from the unmarked clique. A closer examination, however, reveals this is not the case. In \fref{fig:Adjacency-N1024-w512-types}, we plot the probability that the particle will be found at each type of vertex, and the dot-dashed-dashed brown curve is the probability of finding the particle anywhere in the marked clique, i.e., at the $a$, $b$, or $c$ type vertices. We see that this brown curve nearly reaches 1 (actually 0.996, as we will later prove) at time 104.4, which means that nearly all of the probability has moved from the unmarked clique to the marked clique at this time. To keep this probability from escaping marked clique, and to make it build up at the marked $a$ vertex, we propose adding a second stage to the algorithm where the weight is changed from $N/2$ to $1$ (or any value that behaves like the unweighted graph). The result is shown in \fref{fig:Adjacency-twostage-N1024}, where we again plotted the probability in each type of vertex and in the marked clique. The vertical dashed line separates the two stages of the algorithm, with the region left of the line corresponding to the first stage where $w = N/2$, and the right side corresponding to the second stage where $w = 1$. We see that in the first stage, probability has collected at the marked clique, and in the second stage, the probability moves between the marked vertex and the $b$ type vertices, with the success probability reaching 0.996 at time 147.5.

Over the next three sections, we will prove the behaviors exhibited in these figures. In \sref{sec:Laplacian}, we will prove \fref{fig:Laplacian-N1024}, that no matter how strong $w$ gets, the Laplacian quantum walk's behavior is the same as the unweighted graph. In \sref{sec:adjacency}, we will prove \fref{fig:Adjacency-N1024}, that the adjacency quantum walk is no different for most values of $w$, but when $w = N/2$, the success probability is boosted from $1/2$ to 0.820. In \sref{sec:twostage}, we will prove \fref{fig:Adjacency-N1024-w512-types} and \fref{fig:Adjacency-twostage-N1024}, that when $w = N/2$, the adjacency quantum walk builds up a probability of $0.996$ at the marked clique, and then the weight can be changed to $w = 1$ to focus this probability at the marked vertex.

%-------------------------------------------------------------------------------
% Section
%-------------------------------------------------------------------------------

\section{\label{sec:Laplacian}Laplacian Quantum Walk}

In this section, we prove that for large $N$, increasing the weight $w$ of the bridge in the barbell graph does not alter the Laplacian quantum walk search algorithm. That is, we prove the behavior that we saw in \fref{fig:Laplacian-N1024}. To do this, we find the eigenvectors and eigenvalues of $H_L$ \eqref{eq:H-Laplacian-5D} for large $N$ using degenerate perturbation theory \cite{Griffiths2018,Wong5}. This involves expressing the Hamiltonian $H_L$ as its leading-order terms $H_L^{(0)}$ plus higher-order corrections $H_L^{(1)}$, etc.,
\[ H_L = H_L^{(0)} + H_L^{(1)} + \dots, \]
finding the eigenvectors and eigenvalues of $H_L^{(0)}$, and then determining how they change with the addition of $H_L^{(1)}$. Then, we express the initial state \eqref{eq:psi0} as a superposition of the eigenvectors, which makes the state at later time \eqref{eq:psi} straightforward to find. From the state, the success probability \eqref{eq:prob-general} is determined by the norm-square of the amplitude of $\ket{a}$.

We consider three possible scalings of $w$ in the following subsections, and in each case, the relevant eigenvectors and eigenvalues are unchanged by $w$, and hence the evolution and success probability are also asymptotically unchanged by $w$.

%-------------------------------------------------------------------------------

\subsection{\label{subsec:Laplacian-small}Small Weights}

For ``small'' weights, where $w$ scales less than $\sqrt{N}$, i.e., $w = o(\sqrt{N})$, the Hamiltonian's leading- and first-order terms are:
\begin{gather*}
    H_L^{(0)} = -\gamma \begin{pmatrix}
        -\frac{N}{2} + \frac{1}{\gamma} & 0 & 0 & 0 & 0 \\
        0 & 0 & 0 & 0 & 0 \\
        0 & 0 & -\frac{N}{2} & 0 & 0 \\
        0 & 0 & 0 & -\frac{N}{2} & 0 \\
        0 & 0 & 0 & 0 & 0 \\
    \end{pmatrix}, \displaybreak[0] \\
    H_L^{(1)} = -\gamma \begin{pmatrix}
        0 & \sqrt{\frac{N}{2}} & 0 & 0 & 0 \\
        \sqrt{\frac{N}{2}} & 0 & \sqrt{\frac{N}{2}} & 0 & 0 \\
        0 & \sqrt{\frac{N}{2}} & 0 & 0 & 0 \\
        0 & 0 & 0 & 0 & \sqrt{\frac{N}{2}} \\
        0 & 0 & 0 & \sqrt{\frac{N}{2}} & 0 \\
    \end{pmatrix}.
\end{gather*}
That is, $w$ does not contribute to the leading-order terms nor the first-order perturbations. Thus, $w$ will not affect the asymptotic behavior of the algorithm, and the system behaves like the unweighted case, which was analyzed in \cite{Wong7}, and which we summarize now. The eigenvectors and eigenvalues of $H_L^{(0)}$ are
\begin{align}
    &\ket{a}, && \frac{\gamma N}{2} - 1, \nonumber \\
    &\ket{b}, && 0, \vphantom{\frac{N}{2}} \nonumber \\
    &\ket{c}, && \frac{\gamma N}{2}, \label{eq:H-Laplacian-Eigensystem} \\
    &\ket{d}, && \frac{\gamma N}{2}, \nonumber \\
    &\ket{e}, && 0. \vphantom{\frac{N}{2}} \nonumber
\end{align}
Note, $\ket{a}$, $\ket{b}$, and $\ket{e}$ are degenerate when $\gamma$ takes the critical value of
\begin{equation}
    \label{eq:gamma}
    \gamma_{c} = \frac{2}{N}.
\end{equation}
Then, linear combinations of $\ket{a}$, $\ket{b}$, and $\ket{e}$ are eigenvectors of $H_L^{(0)} + H_L^{(1)}$, i.e.,
\begin{align*}
    &\left( H_L^{(0)} + H_L^{(1)} \right) \left( \alpha_a \ket{a} + \alpha_b \ket{b} + \alpha_e \ket{e} \right) \\
    &\quad\quad\quad = E \left( \alpha_a \ket{a} + \alpha_b \ket{b} + \alpha_e \ket{e} \right),
\end{align*}
for some $\alpha_a$, $\alpha_b$, and $\alpha_e$ to be determined next. In matrix form, this eigenvector-eigenvalue relation is
\begin{equation}
    \label{eq:H-Laplacian-Perturbed}
    \begin{pmatrix}
        0 & -\sqrt{\frac{2}{N}} & 0 \\
        -\sqrt{\frac{2}{N}} & 0 & 0 \\
        0 & 0 & 0 \\
    \end{pmatrix} \begin{pmatrix}
        \alpha_a \\
        \alpha_b \\
        \alpha_e \\
    \end{pmatrix} = E \begin{pmatrix}
        \alpha_a \\
        \alpha_b \\
        \alpha_e \\
    \end{pmatrix}.
\end{equation}
Solving this yields the following eigenvectors and eigenvalues of $H_L^{(0)} + H_L^{(1)}$:
\begin{align}
    &\psi_1 = \frac{1}{\sqrt{2}} \left( \ket{a} + \ket{b} \right), && E_1 = -\sqrt{\frac{2}{N}}, \nonumber \\
    &\psi_2 = \frac{1}{\sqrt{2}} \left( -\ket{a} + \ket{b} \right), && E_2 = \sqrt{\frac{2}{N}}, \label{eq:Laplacian-Eigensystem} \\
    &\psi_3 = \ket{e}, && E_3 = 0. \vphantom{\sqrt{\frac{2}{N}}} \nonumber
\end{align}
Now, the initial uniform state \eqref{eq:psi0} can be expressed in terms of these eigenvectors:
\begin{align*}
    \ket{\psi(0)}
        &\approx \frac{1}{\sqrt{2}} \left( \ket{b} + \ket{e} \right) \displaybreak[0] \\
        &= \frac{1}{\sqrt{2}} \left( \frac{1}{\sqrt{2}} \psi_1 + \frac{1}{\sqrt{2}} \psi_2 + \psi_3 \right).
\end{align*}
Following \eqref{eq:psi}, the state at time $t$ is
\begin{align*}
    \ket{\psi(t)} 
        &= e^{-iHt} \ket{\psi(0)} \displaybreak[0] \\
        &\approx e^{-iHt} \frac{1}{\sqrt{2}} \left( \frac{1}{\sqrt{2}} \psi_1 + \frac{1}{\sqrt{2}} \psi_2 + \psi_3 \right) \displaybreak[0] \\
        &= \frac{1}{\sqrt{2}} \left( \frac{1}{\sqrt{2}} e^{-iE_1t} \psi_1 + \frac{1}{\sqrt{2}} e^{-iE_2t} \psi_2 + e^{-iE_3t} \psi_3 \right) \displaybreak[0] \\
        &= \frac{1}{\sqrt{2}} \left[ \frac{1}{2} e^{i\sqrt{2/N}t} \left( \ket{a} + \ket{b} \right) \right. \\
            &\quad\quad\quad+ \left. \frac{1}{2} e^{-i\sqrt{2/N}t} \left( -\ket{a} + \ket{b} \right) + e^0 \ket{e} \right] \displaybreak[0] \\
        &= \frac{1}{\sqrt{2}} \left[ \frac{1}{2} \left( e^{i\sqrt{2/N}t} - e^{-i\sqrt{2/N}t} \right) \ket{a} \right. \\
            &\quad\quad\quad+ \left. \frac{1}{2} \left( e^{i\sqrt{2/N}t} + e^{-i\sqrt{2/N}t} \right) \ket{b} + \ket{e} \right] \displaybreak[0] \\
        &= \frac{1}{\sqrt{2}} \left[ i \sin \left( \sqrt{\frac{2}{N}} t \right) \ket{a} + \cos \left( \sqrt{\frac{2}{N}} t \right) \ket{b} + \ket{e} \right].
\end{align*}
Then taking the norm-square of each amplitude, the probability of measuring the particle at each type of vertex at time $t$ is
\begin{align}
    p_a(t) &= \frac{1}{2} \sin^2 \left( \sqrt{\frac{2}{N}} t \right), \nonumber \displaybreak[0] \\
    p_b(t) &= \frac{1}{2} \cos^2 \left( \sqrt{\frac{2}{N}} t \right), \nonumber \displaybreak[0] \\
    p_c(t) &= 0, \vphantom{\left( \sqrt{\frac{2}{N}} t \right)} \label{eq:prob-unweighted} \displaybreak[0] \\
    p_d(t) &= 0, \vphantom{\left( \sqrt{\frac{2}{N}} t \right)} \nonumber \\
    p_e(t) &= \frac{1}{2}. \nonumber
\end{align}
Note the success probability $p_a(t)$ is consistent with \fref{fig:Laplacian-N1024}.

At the runtime
\begin{equation}
    \label{eq:runtime}
    t_* = \frac{\pi}{2\sqrt{2}} \sqrt{N},
\end{equation}
the success probability reaches its first maximum of
\begin{equation}
    \label{eq:prob-max}
    p_* = p_a(t_*) = \frac{1}{2},
\end{equation}
while $p_b(t_*) = p_c(t_*) = p_d(t_*) = 0$ and $p_e = 1/2$. So, at the runtime, the probability in the marked clique collects at the marked vertex, while the probability in the unmarked clique remains trapped---the unweighted bridge is insufficient to allow the probability to pass from the unmarked clique to the marked clique. Note \eqref{eq:runtime} and \eqref{eq:prob-max} are consistent with \fref{fig:Laplacian-N1024}, where the first peak in success probability occurs at time $\pi\sqrt{1024}/2\sqrt{2} \approx 35.543$ and reaches a height of $1/2$.

Since the success probability reaches $1/2$, on average, one expects to repeat the algorithm twice before finding the marked vertex. Multiplying \eqref{eq:runtime} by two, the expected overall runtime with repetitions is
\begin{equation}
    \label{eq:runtime-total}
    T = \frac{\pi}{\sqrt{2}} \sqrt{N},
\end{equation}
which is approximately $2.221 \sqrt{N}$.

On the other hand, when $\gamma$ is away from its critical value of $\gamma_c = 2/N$ \eqref{eq:gamma}, then an examination of the eigenvectors and eigenvalues of $H_L^{(0)}$ in \eqref{eq:H-Laplacian-Eigensystem} reveals that $\ket{a}$ is no longer degenerate with $\ket{b}$ and $\ket{d}$. As a result, the eigenvectors and eigenvalues of $H_L^{(0)} + H_L^{(1)}$ are no longer those from \eqref{eq:Laplacian-Eigensystem}. Instead, $\ket{b}$ and $\ket{e}$ are each eigenvectors of $H_L^{(0)} + H_L^{(1)}$ with eigenvalue 0, and since the initial state $\ket{\psi(0)} \approx (\ket{b} + \ket{e})/\sqrt{2}$, the system asymptotically stays in this initial state, and the quantum walk does not build up probability at the marked vertex.

%-------------------------------------------------------------------------------

\subsection{\label{subsec:Laplacian-medium}Medium Weights}

Next, when $w$ scales at least as $\sqrt{N}$, i.e., $w = \Omega(\sqrt{N})$, but less than $N$, i.e., $w = o(N)$, then the weight will contribute to $H^{(1)}$. That is,
\begin{gather*}
    H_L^{(0)} = -\gamma \begin{pmatrix}
        -\frac{N}{2} + \frac{1}{\gamma} & 0 & 0 & 0 & 0 \\
        0 & 0 & 0 & 0 & 0 \\
        0 & 0 & -\frac{N}{2} & 0 & 0 \\
        0 & 0 & 0 & -\frac{N}{2} & 0 \\
        0 & 0 & 0 & 0 & 0 \\
    \end{pmatrix}, \displaybreak[0] \\
    H_L^{(1)} = -\gamma \begin{pmatrix}
        0 & \sqrt{\frac{N}{2}} & 0 & 0 & 0 \\
        \sqrt{\frac{N}{2}} & 0 & \sqrt{\frac{N}{2}} & 0 & 0 \\
        0 & \sqrt{\frac{N}{2}} & -w & w & 0 \\
        0 & 0 & w & -w & \sqrt{\frac{N}{2}} \\
        0 & 0 & 0 & \sqrt{\frac{N}{2}} & 0 \\
    \end{pmatrix}.
\end{gather*}
Since $H_L^{(0)}$ is unchanged from the previous case, it has the same eigenvectors and eigenvalues as \eqref{eq:H-Laplacian-Eigensystem}, so $\ket{a}$, $\ket{b}$, and $\ket{e}$ are still degenerate when $\gamma = 2/N$ \eqref{eq:gamma}. Adding the perturbation $H_L^{(1)}$, the weight only affects the states $\ket{c}$ and $\ket{d}$, and so $w$ does not appear when we calculate the eigenvectors of $H_L^{(0)} + H_L^{(1)}$ of the form $\alpha_a \ket{a} + \alpha_b \ket{b} + \alpha_e \ket{e}$. Then, we get the same eigenvalue-eigenvector relation \eqref{eq:H-Laplacian-Perturbed} as before, so we have the same eigenvectors and eigenvalues \eqref{eq:Laplacian-Eigensystem}, and hence the same evolution. Thus, for large $N$, this case with medium weights behaves the same as the case with small weights.

%-------------------------------------------------------------------------------

\subsection{\label{subsec:Laplacian-large}Large Weights}

Finally, when $w$ scales at least as $N$, i.e., $w = \Omega(N)$, the weights will now contribute to $H_L^{(0)}$. That is,
\begin{gather*}
    H_L^{(0)} = -\gamma \begin{pmatrix}
        -\frac{N}{2} + \frac{1}{\gamma} & 0 & 0 & 0 & 0 \\
        0 & 0 & 0 & 0 & 0 \\
        0 & 0 & -\frac{N}{2} - w & w & 0 \\
        0 & 0 & w & -\frac{N}{2} - w & 0 \\
        0 & 0 & 0 & 0 & 0 \\
    \end{pmatrix}, \displaybreak[0] \\
    H_L^{(1)} = -\gamma \begin{pmatrix}
        0 & \sqrt{\frac{N}{2}} & 0 & 0 & 0 \\
        \sqrt{\frac{N}{2}} & 0 & \sqrt{\frac{N}{2}} & 0 & 0 \\
        0 & \sqrt{\frac{N}{2}} & 0 & 0 & 0 \\
        0 & 0 & 0 & 0 & \sqrt{\frac{N}{2}} \\
        0 & 0 & 0 & \sqrt{\frac{N}{2}} & 0 \\
    \end{pmatrix}.
\end{gather*}
Note that $H_L^{(0)} + H_L^{(1)}$ is identical to the ``medium weight'' case in the previous subsection, so it has the same eigenvectors, so the weight does not play a role.

Alternatively, if one proceeds with the perturbative calculation, the eigenvectors and eigenvalues of $H_L^{(0)}$ are
\begin{align*}
    &\ket{a}, \vphantom{\frac{1}{\sqrt{2}}} && \frac{\gamma N}{2} - 1, \displaybreak[0] \\
    &\ket{b}, \vphantom{\frac{1}{\sqrt{2}}} && 0, \displaybreak[0] \\
    &\ket{cd} = \frac{1}{\sqrt{2}} \left( \ket{c} + \ket{d} \right), && \frac{\gamma N}{2}, \displaybreak[0] \\
    &\ket{cd^-} = \frac{1}{\sqrt{2}} \left( \ket{c} - \ket{d} \right), && \frac{\gamma N}{2} + 2\gamma w, \\
    &\ket{e}, \vphantom{\frac{1}{\sqrt{2}}} && 0.
\end{align*}
As in the other cases, $\ket{a}$, $\ket{b}$, and $\ket{e}$ are degenerate when $\gamma = 2/N$ \eqref{eq:gamma}. Adding the perturbation, the eigenvectors of $H_L^{(0)} + H_L^{(1)}$ that are linear combinations of $\ket{a}$, $\ket{b}$, and $\ket{e}$ satisfy the same eigenvalue-eigenvector relation \eqref{eq:H-Laplacian-Perturbed} as the previous cases, resulting in the same eigenvectors and eigenvalues \eqref{eq:Laplacian-Eigensystem}, and the same asymptotic evolution.

%-------------------------------------------------------------------------------
% Section
%-------------------------------------------------------------------------------

\section{\label{sec:adjacency}Adjacency Quantum Walk}

In this section, we prove the behavior that we saw in \fref{fig:Adjacency-N1024} and save the behaviors that we saw in \fref{fig:Adjacency-N1024-w512-types} and \fref{fig:Adjacency-twostage-N1024} for the next section. As in the previous section, we express the Hamiltonian 
\eqref{eq:H-Adjacency-5D} as leading- and higher-order terms
\[ H_A = H_A^{(0)} + H_A^{(1)} + \dots, \]
and employ degenerate perturbation theory to find the eigenvectors and eigenvalues of the Hamiltonian for large $N$. We show that most values of $w$ do not change the relevant eigenvectors and eigenvalues, but when $w = 1/\gamma = N/2$, the evolution is qualitatively and quantitatively different.

%-------------------------------------------------------------------------------

\subsection{\label{subsec:adjacency-small}Small and Medium Weights}

As discussed in the second-to-last paragraph of the Introduction, when $w$ scales less than $N$, the weighted barbell graph is approximately regular with degree $N/2$. Then, the Hamiltonians for the Laplacian and adjacency quantum walks asymptotically differ by a multiple of the identity matrix, namely $H_L = H_A + \gamma D$ with $D = (N/2)I$, and any multiple of the identity matrix can be dropped as a rezeroing of energy. Alternatively, the multiple of the identity matrix manifests itself as a global phase, which can be dropped. Thus, the Laplacian and adjacency quantum walks behave the same way when $w$ scales less than $N$.

To see this explicitly using degenerate perturbation theory, for small weights where $w = o(\sqrt{N})$, the leading- and first-order terms of $H_A$ are
\begin{gather*}
    H_A^{(0)} = -\gamma \begin{pmatrix}
        \frac{1}{\gamma} & 0 & 0 & 0 & 0 \\
        0 & \frac{N}{2} & 0 & 0 & 0 \\
	    0 & 0 & 0 & 0 & 0 \\
        0 & 0 & 0 & 0 & 0 \\
        0 & 0 & 0 & 0 & \frac{N}{2} \\
    \end{pmatrix}, \displaybreak[0] \\
    H_A^{(1)} = -\gamma \begin{pmatrix}
        0 & \sqrt{\frac{N}{2}} & 0 & 0 & 0 \\
        \sqrt{\frac{N}{2}} & 0 & \sqrt{\frac{N}{2}} & 0 & 0 \\
	    0 & \sqrt{\frac{N}{2}} & 0 & 0 & 0 \\
        0 & 0 & 0 & 0 & \sqrt{\frac{N}{2}} \\
        0 & 0 & 0 & \sqrt{\frac{N}{2}} & 0 \\
    \end{pmatrix}.
\end{gather*}
Comparing these with $H_L^{(0)}$ and $H_L^{(1)}$ in \sref{subsec:Laplacian-small}, $H_A^{(0)} = H_L^{(0)} - (\gamma N/2)I$, and $H_A^{(1)} = H_L^{(1)}$. Then, $H_A^{(0)}$ has the same eigenvectors as $H_L^{(0)}$, and the corresponding eigenvalues of $H_A^{(0)}$ can be obtained by subtracting $\gamma N/2$ from the eigenvalues of $H_L^{(0)}$. The same eigenvectors are degenerate when $\gamma = 2/N$. When the perturbation is included, since $H_A^{(1)} = H_L^{(1)}$, the eigenvectors of $H_A^{(0)} + H_A^{(1)}$ are the same as $H_L^{(0)} + H_L^{(1)}$, and the eigenvalues are also shifted by $-\gamma N/2 = -1$. Expressing the initial state \eqref{eq:psi0} in terms of these eigenvectors and evolving the system \eqref{eq:psi}, the shift in the eigenvalue of $-1$ manifests itself as a global phase $e^{it}$, which can be dropped, or equivalently disappears when finding the success probability \eqref{eq:prob-general}, so the success probability also evolves as \eqref{eq:prob-unweighted}.

Similarly, for medium weights where $w = \Omega(\sqrt{N})$ and $w = o(N)$,
\begin{gather*}
    H_A^{(0)} = -\gamma \begin{pmatrix}
        \frac{1}{\gamma} & 0 & 0 & 0 & 0 \\
        0 & \frac{N}{2} & 0 & 0 & 0 \\
	    0 & 0 & 0 & 0 & 0 \\
        0 & 0 & 0 & 0 & 0 \\
        0 & 0 & 0 & 0 & \frac{N}{2} \\
    \end{pmatrix}, \displaybreak[0] \\
    H_A^{(1)} = -\gamma \begin{pmatrix}
        0 & \sqrt{\frac{N}{2}} & 0 & 0 & 0 \\
        \sqrt{\frac{N}{2}} & 0 & \sqrt{\frac{N}{2}} & 0 & 0 \\
	    0 & \sqrt{\frac{N}{2}} & -w & w & 0 \\
        0 & 0 & w & -w & \sqrt{\frac{N}{2}} \\
        0 & 0 & 0 & \sqrt{\frac{N}{2}} & 0 \\
    \end{pmatrix}.
\end{gather*}
Comparing these with $H_L^{(0)}$ and $H_L^{(1)}$ in \sref{subsec:Laplacian-medium}, we again have $H_A^{(0)} = H_L^{(0)} - (\gamma N/2)I$, and $H_A^{(1)} = H_L^{(1)}$, and so the evolutions are the identical by the same argument as above.

%-------------------------------------------------------------------------------

\subsection{Large Weights, Part 1}

For the Laplacian quantum walk, we considered when $w$ scales as $N$ or greater than $N$ together in \sref{subsec:Laplacian-large}. For the adjacency quantum walk, however, we consider these separately, with $w$ scaling as $N$ here and $w$ scaling greater than $N$ in the next subsection.

When $w = \Theta(N)$,
\begin{equation}
\label{eq:adjacency-large1-H0H1}
\begin{gathered}
    H_A^{(0)} = -\gamma \begin{pmatrix}
        \frac{1}{\gamma} & 0 & 0 & 0 & 0 \\
        0 & \frac{N}{2} & 0 & 0 & 0 \\
        0 & 0 & 0 & w & 0 \\
        0 & 0 & w & 0 & 0 \\
        0 & 0 & 0 & 0 & \frac{N}{2} \\    
    \end{pmatrix}, \\
    H_A^{(1)} = -\gamma \begin{pmatrix} 
        0 & \sqrt{\frac{N}{2}} & 0 & 0 & 0 \\
        \sqrt{\frac{N}{2}} & 0 & \sqrt{\frac{N}{2}} & 0 & 0 \\
        0 & \sqrt{\frac{N}{2}} & 0 & 0 & 0 \\
        0 & 0 & 0 & 0 & \sqrt{\frac{N}{2}} \\
        0 & 0 & 0 & \sqrt{\frac{N}{2}} & 0 \\
    \end{pmatrix}.
\end{gathered}
\end{equation}
The eigenvectors and eigenvalues of $H_A^{(0)}$ are
\begin{align}
    &\ket{a}, \vphantom{\frac{1}{\sqrt{2}}} && {-1}, \displaybreak[0] \nonumber \\
    &\ket{b}, \vphantom{\frac{1}{\sqrt{2}}} && \frac{-\gamma N}{2} \displaybreak[0] \nonumber \\
    &\ket{cd} = \frac{1}{\sqrt{2}} \left( \ket{c} + \ket{d} \right), && {-\gamma w}, \displaybreak[0] \label{eq:adjacency-large1-H0-eigensystem} \\
    &\ket{cd^-} = \frac{1}{\sqrt{2}} \left( \ket{c} - \ket{d} \right), && \gamma w, \displaybreak[0] \nonumber \\
    &\ket{e}, \vphantom{\frac{1}{\sqrt{2}}} && \frac{-\gamma N}{2}. \nonumber
\end{align}
Comparing the above eigenvalues, $\ket{b}$ and $\ket{e}$ are always degenerate; $\ket{a}$, $\ket{b}$, and $\ket{e}$ are degenerate when $\gamma = 2/N$; $\ket{b}$, $\ket{cd}$, and $\ket{e}$ are degenerate when $w = 1/\gamma$; and $\ket{a}$, $\ket{b}$, $\ket{cd}$; and $\ket{e}$ are degenerate when both $\gamma = 2/N$ and $w = 1/\gamma = N/2$.

\begin{figure}
\begin{center}
    \includegraphics{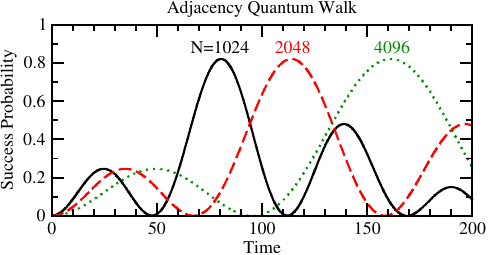}
    \caption{\label{fig:Adjacency-Nvaries}Success probability versus time for search on the barbell of graph of $N$ vertices using an adjacency quantum walk with $\gamma = 2/N$ and $w = 1/\gamma = N/2$. The solid black curve is $N = 1024$, the dashed red curve is $N = 2048$, and the dotted green curve is $N = 4096$.}
\end{center}
\end{figure}

Let us consider the last case, when $\gamma = 2/N$ and $w = 1/\gamma = N/2$, so $\ket{a}$, $\ket{b}$, $\ket{cd}$, and $\ket{e}$ are degenerate eigenvectors of $H_A^{(0)}$. Then, linear combinations of them,
\[ \ket{\psi} = \alpha_a \ket{a} + \alpha_b \ket{b} + \alpha_{cd} \ket{cd} + \alpha_e \ket{e}, \]
are eigenvectors of $H_A^{(0)} + H_A^{(1)}$. The $\alpha$ coefficients of these eigenvectors, and the eigenvalues $E$, can be found by solving the eigenvalue-eigenvector relation
\[ \left( H_A^{(0)} + H_A^{(1)} \right) \ket{\psi} = E \ket{\psi}, \]
which in matrix form is
\[ \begin{pmatrix}
    -1 & -\sqrt{\frac{2}{N}} & 0 & 0 \\
    -\sqrt{\frac{2}{N}} & -1 & -\frac{1}{\sqrt{N}} & 0 \\
    0 & -\frac{1}{\sqrt{N}} & -1 & -\frac{1}{\sqrt{N}} \\
    0 & 0 & -\frac{1}{\sqrt{N}} & -1 \\
\end{pmatrix} \begin{pmatrix}
    \alpha_a \\
    \alpha_b \\
    \alpha_{cd} \\
    \alpha_{e} \\
\end{pmatrix} = E \begin{pmatrix}
    \alpha_a \\
    \alpha_b \\
    \alpha_{cd} \\
    \alpha_{e} \\
\end{pmatrix}. \]
Solving this, the (unnormalized) eigenvectors of $H_A^{(0)} + H_A^{(1)}$ are
\begin{align*}
    &\psi_1 = \sqrt{2 + \sqrt{2}} \ket{a} + \left( 1 + \sqrt{2} \right) \ket{b} + \sqrt{2 + \sqrt{2}} \ket{cd} + \ket{e}, \displaybreak[0] \\
    &\psi_2 = -\sqrt{2 + \sqrt{2}} \ket{a} + \left( 1 + \sqrt{2} \right) \ket{b} - \sqrt{2 + \sqrt{2}} \ket{cd} + \ket{e}, \displaybreak[0] \\
    &\psi_3 = -\sqrt{2 - \sqrt{2}} \ket{a} + \left( 1 - \sqrt{2} \right) \ket{b} + \sqrt{2 - \sqrt{2}} \ket{cd} + \ket{e}, \displaybreak[0] \\
    &\psi_4 = \sqrt{2 - \sqrt{2}} \ket{a} + \left( 1 - \sqrt{2} \right) \ket{b} - \sqrt{2 - \sqrt{2}} \ket{cd} + \ket{e},
\end{align*}
with corresponding eigenvalues
\begin{align*}
    &E_1 = -1 - \sqrt{\frac{2 + \sqrt{2}}{N}}, \displaybreak[0] \\
    &E_2 = -1 + \sqrt{\frac{2 + \sqrt{2}}{N}}, \displaybreak[0] \\
    &E_3 = -1 - \sqrt{\frac{2 - \sqrt{2}}{N}}, \displaybreak[0] \\
    &E_4 = -1 + \sqrt{\frac{2 - \sqrt{2}}{N}}.
\end{align*}
Now, we can express the initial state \eqref{eq:psi0} as a linear combination of the eigenvectors of $H_A^{(0)} + H_A^{(1)}$:
\begin{align*}
    \ket{\psi(0)}
        &\approx \frac{1}{\sqrt{2}} \left( \ket{b} + \ket{e} \right) \displaybreak[0] \\
        &= \frac{1}{4\sqrt{2}} \left( \psi_1 + \psi_2 + \psi_3 + \psi_4 \right).
\end{align*}
Then, the state at time $t$ \eqref{eq:psi} is
\begin{widetext}
\begin{align}
    \ket{\psi(t)} 
        &= e^{-iHt} \ket{\psi(0)} \nonumber \nonumber \displaybreak[0] \\
        &\approx e^{-iHt} \frac{1}{4\sqrt{2}} \left(\psi_1 + \psi_2 + \psi_3 + \psi_4 \right) \nonumber \displaybreak[0] \\
        &= \frac{1}{4\sqrt{2}} \Big( e^{-iE_1t} \psi_1 + e^{-iE_2t} \psi_2 + e^{-iE_3t} \psi_3 + e^{-iE_4t} \psi_4 \Big) \nonumber \displaybreak[0] \\
        &= \frac{1}{4\sqrt{2}} \left\{ e^{-i \left(-1 - \sqrt{\frac{2+\sqrt{2}}{N}} \right) t} \left[ \sqrt{2 + \sqrt{2}} \ket{a} + \left( 1 + \sqrt{2} \right) \ket{b} + \sqrt{2 + \sqrt{2}} \ket{cd} + \ket{e} \right] \right. \nonumber \\
            &\quad\quad\quad\quad + e^{-i \left( -1 + \sqrt{\frac{2 + \sqrt{2}}{N}} \right) t} \left[ -\sqrt{2 + \sqrt{2}} \ket{a} + \left( 1 + \sqrt{2} \right) \ket{b} - \sqrt{2 + \sqrt{2}} \ket{cd} + \ket{e} \right] \nonumber \\
            &\quad\quad\quad\quad + e^{-i \left( -1 - \sqrt{\frac{2 - \sqrt{2}}{N}} \right) t} \left[ -\sqrt{2 - \sqrt{2}} \ket{a} + \left( 1 - \sqrt{2} \right) \ket{b} + \sqrt{2 - \sqrt{2}} \ket{cd} + \ket{e} \right] \nonumber \\
            &\quad\quad\quad\quad + \left. e^{-i \left( -1 + \sqrt{\frac{2 - \sqrt{2}}{N}} \right) t} \left[ \sqrt{2 - \sqrt{2}} \ket{a} + \left( 1 - \sqrt{2} \right) \ket{b} - \sqrt{2 - \sqrt{2}} \ket{cd} + \ket{e} \right] \right\} \nonumber \displaybreak[0] \\
        &= \frac{e^{it}}{4\sqrt{2}} \left\{ \left[ \left( e^{i \sqrt{\frac{2+\sqrt{2}}{N}} t} - e^{-i \sqrt{\frac{2+\sqrt{2}}{N}} t} \right) \sqrt{2+\sqrt{2}} - \left( e^{i \sqrt{\frac{2-\sqrt{2}}{N}} t} - e^{-i \sqrt{\frac{2-\sqrt{2}}{N}} t} \right) \sqrt{2-\sqrt{2}} \ \right] \ket{a} \right. \nonumber \\
            &\quad\quad\quad\quad + \left[ \left( e^{i \sqrt{\frac{2+\sqrt{2}}{N}} t} + e^{-i \sqrt{\frac{2 + \sqrt{2}}{N}} t} \right) \left( 1 + \sqrt{2} \right) + \left( e^{i \sqrt{\frac{2-\sqrt{2}}{N}} t} + e^{-i \sqrt{\frac{2-\sqrt{2}}{N}} t} \right) \left( 1 - \sqrt{2} \right) \right] \ket{b} \nonumber \\
            &\quad\quad\quad\quad + \left[ \left( e^{i \sqrt{\frac{2+\sqrt{2}}{N}} t} - e^{-i \sqrt{\frac{2+\sqrt{2}}{N}} t} \right) \sqrt{2+\sqrt{2}} + \left( e^{i \sqrt{\frac{2-\sqrt{2}}{N}} t} - e^{-i \sqrt{\frac{2-\sqrt{2}}{N}} t} \right) \sqrt{2-\sqrt{2}} \ \right] \ket{cd} \nonumber \\
            &\quad\quad\quad\quad + \left. \left[ \left( e^{i \sqrt{\frac{2+\sqrt{2}}{N}} t} + e^{-i \sqrt{\frac{2 + \sqrt{2}}{N}} t} \right) + \left( e^{i \sqrt{\frac{2-\sqrt{2}}{N}} t} + e^{-i \sqrt{\frac{2-\sqrt{2}}{N}} t} \right) \right] \ket{e} \right\} \nonumber \displaybreak[0] \\
        &= \frac{e^{it}}{4\sqrt{2}} \left\{ 2i \left[ \sqrt{2+\sqrt{2}}  \sin \left( \sqrt{\frac{2+\sqrt{2}}{N}} t \right) - \sqrt{2-\sqrt{2}} \sin \left( \sqrt{\frac{2-\sqrt{2}}{N}} t \right) \right] \ket{a} \right. \nonumber \\
            &\quad\quad\quad\quad + 2 \left[ \left( 1 + \sqrt{2} \right) \cos \left( \sqrt{\frac{2+\sqrt{2}}{N}} t \right) + \left( 1 - \sqrt{2} \right) \cos \left( \sqrt{\frac{2-\sqrt{2}}{N}} t \right) \right] \ket{b} \nonumber \\
            &\quad\quad\quad\quad + 2i \left[ \sqrt{2+\sqrt{2}}  \sin \left( \sqrt{\frac{2+\sqrt{2}}{N}} t \right) + \sqrt{2-\sqrt{2}} \sin \left( \sqrt{\frac{2-\sqrt{2}}{N}} t \right) \right] \ket{cd} \nonumber \\
            &\quad\quad\quad\quad + \left. 2 \left[ \cos \left( \sqrt{\frac{2+\sqrt{2}}{N}} t \right) + \cos \left( \sqrt{\frac{2-\sqrt{2}}{N}} t \right) \right] \ket{e} \right\}. \label{eq:adjacency-large1-amplitudes}
\end{align}
Using $\ket{cd} = \left( \ket{c} + \ket{d} \right) / \sqrt{2}$ and taking the norm-square of each amplitude, the probability of measuring the particle to be at each type of vertex is
\begin{align}
    p_a(t) &= \frac{1}{8} \left[ \sqrt{2+\sqrt{2}} \sin \left( \sqrt{\frac{2+\sqrt{2}}{N}} t \right) - \sqrt{2-\sqrt{2}} \sin \left( \sqrt{\frac{2-\sqrt{2}}{N}} t \right) \right]^2, \nonumber \displaybreak[0] \\
    p_b(t) &= \frac{1}{8} \left[ \left( 1 + \sqrt{2} \right) \cos \left( \sqrt{\frac{2+\sqrt{2}}{N}} t \right) + \left( 1 - \sqrt{2} \right) \cos \left( \sqrt{\frac{2-\sqrt{2}}{N}} t \right) \right]^2, \nonumber \displaybreak[0] \\
    p_c(t) &= \frac{1}{16} \left[ \sqrt{2+\sqrt{2}}  \sin \left( \sqrt{\frac{2+\sqrt{2}}{N}} t \right) + \sqrt{2-\sqrt{2}} \sin \left( \sqrt{\frac{2-\sqrt{2}}{N}} t \right) \right]^2, \label{eq:adjacency-large1-probabilities} \displaybreak[0] \\
    p_d(t) &= \frac{1}{16} \left[ \sqrt{2+\sqrt{2}}  \sin \left( \sqrt{\frac{2+\sqrt{2}}{N}} t \right) + \sqrt{2-\sqrt{2}} \sin \left( \sqrt{\frac{2-\sqrt{2}}{N}} t \right) \right]^2, \nonumber \displaybreak[0] \\
    p_e(t) &= \frac{1}{8} \left[ \cos \left( \sqrt{\frac{2+\sqrt{2}}{N}} t \right) + \cos \left( \sqrt{\frac{2-\sqrt{2}}{N}} t \right) \right]^2. \nonumber
\end{align}
Note the success probability $p_a(t)$ reproduces the dotted green curve in \fref{fig:Adjacency-N1024} when $N = 1024$.

To prove that the success probability reaches a peak of 0.820 and find the corresponding runtime, we take the derivative of $p_a(t)$ and set it equal to zero. Its derivative is
\begin{align*}
    \frac{dp_a}{dt} 
        &= \frac{1}{4} \left[ \sqrt{2-\sqrt{2}} \sin \left( \frac{\sqrt{2-\sqrt{2}}}{\sqrt{N}} t \right) - \sqrt{2+\sqrt{2}} \sin \left( \frac{\sqrt{2+\sqrt{2}}}{\sqrt{N}} t \right) \right] \\
        &\quad \times \left[ \frac{2-\sqrt{2}}{\sqrt{N}} \cos \left( \frac{\sqrt{2-\sqrt{2}}}{\sqrt{N}} t \right) - \frac{2+\sqrt{2}}{\sqrt{N}} \cos \left( \frac{\sqrt{2+\sqrt{2}}}{\sqrt{N}} t \right) \right].
\end{align*}
Then we set $dp_a/dt$ equal to zero to find the minima and maxima. For this problem, the relevant solution set comes from the cosine portion, so we set it equal to zero and multiply by $\sqrt{N}$ to simplify, resulting in
\[ \left( 2-\sqrt{2} \right) \cos \left( \frac{\sqrt{2-\sqrt{2}}}{\sqrt{N}} t \right) - \left( 2+\sqrt{2} \right) \cos \left( \frac{\sqrt{2+\sqrt{2}}}{\sqrt{N}} t \right) = 0. \]
\end{widetext}
This is a transcendental equation for $t/\sqrt{N}$, so it does not have an analytical solution, but it can be solved numerically such that peak probability is reached when $t/\sqrt{N} \approx 2.518$, or
\begin{equation}
    \label{eq:adjacency-large1-runtime}
    t_* \approx 2.518 \sqrt{N}.
\end{equation}
When plugging this value of $t$ into our success probability \eqref{eq:adjacency-large1-probabilities}, the $\sqrt{N}$'s cancel, and we get
\begin{equation}
    \label{eq:adjacency-large1-prob}
    p_* = p_a(t_*) \approx 0.820.
\end{equation}
Note that this peak does not depend on $N$. As a check, we have plotted in \fref{fig:Adjacency-Nvaries} the success probability versus time of the search algorithm with $w = N/2$ for various values of $N$. We see that the peak occurs at 0.820 in each of the cases, confirming that this value does not depend on $N$. This figure also confirms the runtime expression \eqref{eq:adjacency-large1-runtime}, since the solid black curve peaks at time $2.518 \sqrt{1024} \approx 80.6$ (as also seen in \fref{fig:Adjacency-N1024}), the dashed red curve peaks at time $2.518 \sqrt{2048} \approx 114.0$, and the dotted green curve peaks at time $2.518 \sqrt{4096} \approx 161.2$.

Comparing the unweighted barbell graph (and the weighted graphs that behave like it) to the case where $w = 1/\gamma = N/2$, we see that the unweighted graph had a success probability that reached $1/2$ \eqref{eq:prob-max} at time $\pi\sqrt{N}/2\sqrt{2}$ \eqref{eq:runtime}, and so the expected total runtime with repetitions was $\pi\sqrt{N}/\sqrt{2} \approx 2.221 \sqrt{N}$ \eqref{eq:runtime-total}. Now, we have shown that when $w = 1/\gamma = N/2$, the success probability is improved to $0.820$ \eqref{eq:adjacency-large1-prob}, and the time at which this occurs is $2.518 \sqrt{N}$ \eqref{eq:adjacency-large1-runtime}. This single runtime of $2.518 \sqrt{N}$, which does not include repetitions, is longer runtime than the unweighted graph's expected runtime of $2.221 \sqrt{N}$ that does include repetitions. Thus, when repetitions are permitted, searching on an unweighted graph may be a faster approach, but if repetitions are not permitted, then searching on the weighted graph for a longer amount of time in order to reach a higher success probability may be better.

\begin{figure}
\begin{center}
    \subfloat[] {
        \includegraphics{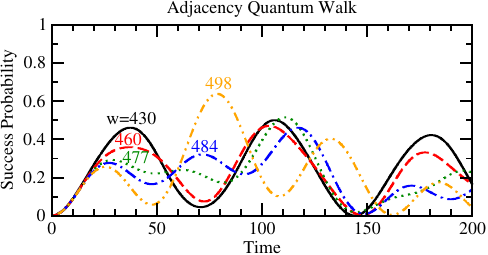}
        \label{fig:Adjacency-N1024-Transition-a}
    }
    
    \subfloat[] {
        \includegraphics{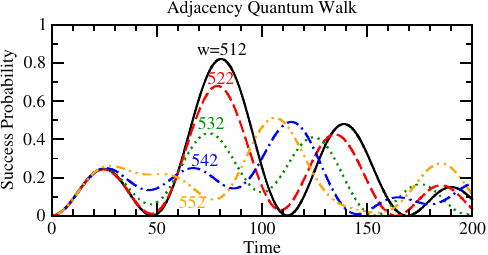}
        \label{fig:Adjacency-N1024-Transition-b}
    }
    
    \subfloat[] {
        \includegraphics{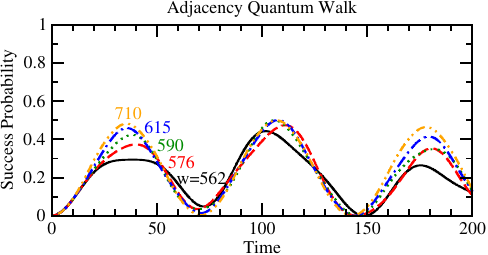}
        \label{fig:Adjacency-N1024-Transition-c}
    }
    \caption{\label{fig:Adjacency-N1024-Transition}Success probability versus time for search on the barbell of graph of $N = 1024$ vertices using an adjacency quantum walk with $\gamma = 2/N$. In (a), the solid black curve is $w = 430$, the dashed red curve is $w = 460$, the dotted green curve is $w = 477$, the dot-dashed blue curve is $w = 484$, and the dot-dot-dashed orange curve is $w = 498$. In (b), the curves are respectively $w = 512$, $522$, $532$, $542$, and $552$. In (c), the curves are respectively $w = 562$, $576$, $590$, $615$, and $710$.}
\end{center}
\end{figure}

In \fref{fig:Adjacency-N1024}, the values of $w$ took large steps. We expect that with smaller steps, there will be a smooth transition to and from the $w = 1/\gamma = N/2$ case that was just analyzed. We show this transition in \fref{fig:Adjacency-N1024-Transition}, and in \fref{fig:Adjacency-N1024-Transition-a}, as $w$ increases, the first local maximum in success probability begins to dip, while a second maximum begins to rise. In \fref{fig:Adjacency-N1024-Transition-b}, the second maximum peaks at a height of $0.820$ when $w = N/2$, and for weights beyond this, it peak diminishes and disappears. In \fref{fig:Adjacency-N1024-Transition-c}, as $w$ continues to increase, the first peak rises again, and the evolution returns to its quasi-periodic form.

%-------------------------------------------------------------------------------

\subsection{Large Weights, Part 2}

When $w = \Omega(N)$, $H_A^{(0)}$ and $H_A^{(1)}$ are identical to the previous case where $w = \Theta(N)$, so the matrices are the same as \eqref{eq:adjacency-large1-H0H1}, and the eigenvectors and eigenvalues of $H_A^{(1)}$ are the same as \eqref{eq:adjacency-large1-H0-eigensystem}. Now, since $w = \Omega(N)$ it must be that $w \ne N/2$, and so $\ket{cd}$ is never degenerate with the other eigenvectors. Thus, when $\gamma = N/2$, only $\ket{a}$, $\ket{b}$, and $\ket{e}$ are degenerate with eigenvalue $-1$. Linear combinations of these,
\[ \alpha_a \ket{a} + \alpha_b \ket{b} + \alpha_c \ket{c}, \]
are eigenvectors of the perturbed Hamiltonian $H_A^{(0)} + H_A^{(1)}$ with eigenvalues $E$, satisfying
\[ \begin{pmatrix}
    -1 & -\sqrt{\frac{2}{N}} & 0 \\
    -\sqrt{\frac{2}{N}} & -1 & 0 \\
    0 & 0 & -1 \\
\end{pmatrix} \begin{pmatrix}
    \alpha_a \\
    \alpha_b \\
    \alpha_e \\
\end{pmatrix} = E  \begin{pmatrix}
    \alpha_a \\
    \alpha_b \\
    \alpha_e \\
\end{pmatrix}. \]
Solving this, three eigenvectors of $H_A^{(0)} + H_A^{(1)}$ are
\begin{align*}
    &\psi_1 = \frac{1}{\sqrt{2}} \left( \ket{a} + \ket{b} \right), && E_{ab} = -1 - \sqrt{\frac{2}{N}}, \displaybreak[0] \\
    &\psi_2 = \frac{1}{\sqrt{2}} \left( \ket{a} + \ket{b} \right), && E_{ab^-} = -1 + \sqrt{\frac{2}{N}}, \displaybreak[0] \\
    &\psi_3 = \ket{e}, && E_e = -1. \vphantom{\sqrt{\frac{2}{N}}}
\end{align*}
Note these are the same eigenvectors and eigenvalues as the Laplacian quantum walk \eqref{eq:Laplacian-Eigensystem}, except the eigenvalues are each shifted by $-1$. Then, the evolution is the same as the Laplacian quantum walk's multiplied by a global phase of $e^{it}$, and so the success probability evolves the same way, i.e., as \eqref{eq:prob-unweighted}.

\iffalse
\begin{gather*}
    H_A^{(0)} = -\gamma \begin{pmatrix}
        \frac{1}{\gamma} & 0 & 0 & 0 & 0 \\
        0 & \frac{N}{2} & 0 & 0 & 0 \\
	    0 & 0 & 0 & w & 0 \\
        0 & 0 & w & 0 & 0 \\
        0 & 0 & 0 & 0 & \frac{N}{2} \\
    \end{pmatrix}, \displaybreak[0] \\
    H_A^{(1)} = -\gamma \begin{pmatrix}
        0 & \sqrt{\frac{N}{2}} & 0 & 0 & 0 \\
        \sqrt{\frac{N}{2}} & 0 & \sqrt{\frac{N}{2}} & 0 & 0 \\
	    0 & \sqrt{\frac{N}{2}} & 0 & 0 & 0 \\
        0 & 0 & 0 & 0 & \sqrt{\frac{N}{2}} \\
        0 & 0 & 0 & \sqrt{\frac{N}{2}} & 0 \\
    \end{pmatrix}.
\end{gather*}
The eigenvectors and eigenvalues of $H_A^{(0)}$ are
\begin{align*}
    &\ket{a}, && {-1}, \displaybreak[0] \\
    &\ket{b} && \frac{-\gamma N}{2}, \displaybreak[0] \\
    &\frac{1}{\sqrt{2}} \left( \ket{c} + \ket{d} \right), && {-\gamma w}, \displaybreak[0] \\
    &\frac{1}{\sqrt{2}} \left( \ket{c} - \ket{d} \right), && \gamma w, \\
    &\ket{e} && \frac{-\gamma N}{2}.
\end{align*}

When $\gamma = 2/N$, 
\fi

%-------------------------------------------------------------------------------
% Section
%-------------------------------------------------------------------------------

\section{\label{sec:twostage}Two-Stage Algorithm}

In this section, we prove the behavior of the two-stage algorithm that was introduced in \fref{fig:Adjacency-twostage-N1024}, which uses the adjacency quantum walk with $w = 1/\gamma = N/2$ for the first stage, and then $w = 1$ (or other values that evolve similarly) for the second stage.

For the first stage, we evolve until the probability in the marked clique peaks near 1. For large $N$, the probability in the marked clique can be obtained by adding $p_a$, $p_b$, and $p_c$ from \eqref{eq:adjacency-large1-probabilities}, i.e.,
\[ p_{abc}(t) = p_a(t) + p_b(t) + p_c(t). \]
This expression agrees with the entire dot-dashed-dashed brown curve in \fref{fig:Adjacency-N1024-w512-types}, and the same curve in \fref{fig:Adjacency-twostage-N1024} during its first stage. Now, to find where this probability peaks (at which we switch to the second stage of the algorithm), we take its derivative and set it equal to zero. Its derivative simplifies to
\begin{align*}
    \frac{dp_{abc}}{dt} 
        &= \frac{1}{4\sqrt{2N}} \left[ \cos \left( \sqrt{\frac{2-\sqrt{2}}{N}} t \right) - \cos \left( \sqrt{\frac{2+\sqrt{2}}{N}} t \right) \right] \\
        &\quad\quad\quad\; \times \left[ \sqrt{2-\sqrt{2}} \sin \left( \sqrt{\frac{2-\sqrt{2}}{N}} t \right) \right. \\
        &\quad\quad\quad\quad\quad + \left. \sqrt{\sqrt{2}+2} \sin \left( \sqrt{\frac{2+\sqrt{2}}{N}} t \right) \right].
\end{align*}
Setting this equal to zero to find the extrema, the relevant solution comes from the sine portion, so we solve
\begin{align*}
    &\sqrt{2-\sqrt{2}} \sin \left( \sqrt{\frac{2-\sqrt{2}}{N}} t \right) \\
        &\quad\quad + \sqrt{\sqrt{2}+2} \sin \left( \sqrt{\frac{2+\sqrt{2}}{N}} t \right) = 0.
\end{align*}
This is a transcendental equation for $t/\sqrt{N}$. Solving it numerically, we get $t/\sqrt{N} \approx 3.265$, or a first stage runtime of
\begin{equation}
    \label{eq:stage1-runtime}
    t' = 3.265 \sqrt{N}.
\end{equation}
Plugging this into $p_{abc}$, we find that
\begin{equation}
    \label{eq:stage1-prob}
    p_{abc}(t') = 0.996,
\end{equation}
so nearly all of the probability is in the marked clique after the first stage. The state at this time is obtained from \eqref{eq:adjacency-large1-amplitudes}, and dropping the irrelevant global phase of $e^{it}$, we get
\[ \ket{\psi(t')} = -0.324 i \ket{a} + 0.944 \ket{b} + 0.060 \ket{e}. \]

\begin{figure*}
\begin{center}
    \includegraphics{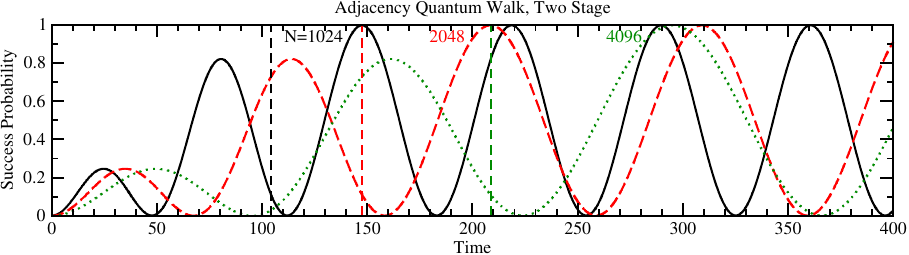}
    \caption{\label{fig:Adjacency-twostage-Nvaries}Success probability versus time for search on the barbell of graph of $N$ vertices using an adjacency quantum walk with $\gamma = 2/N$. The solid black curve is $N = 1024$, the dashed red curve is $N = 2048$, and the dotted green curve is $N = 4096$. The dashed vertical lines separate the two stages for each of the curves; left of the vertical line, $w = N/2$, and right of the line, $w = 1$.}
\end{center}
\end{figure*}

The above state is the also state of the system at the beginning of the second stage of the algorithm. Now, we let $w = 1$. As we analyzed in \sref{subsec:adjacency-small}, the eigenvectors and eigenvalues of $H_A^{(0)} + H_A^{(1)}$ are asymptotically identical to Laplacian quantum walks from \eqref{eq:Laplacian-Eigensystem}, except $-1$ is added to each eigenvalue, which constitutes an irrelevant rezeroing of energy or global phase. Expressing $\ket{\psi(t_1)}$ as a linear combination of the eigenvectors,
\begin{align*}
    \ket{\psi(t')} 
        &= 0.706 e^{-0.331 i} \psi_1 + 0.706 e^{-0.331 i} \psi_2 \\
        &\quad + 0.060 \psi_3.
\end{align*}
Let $\Delta t = t - t'$ denote the additional time that the algorithm runs during its second stage. Then, the system at time $t > t'$ is
\begin{widetext}
\begin{align*}
    \ket{\psi(t)} 
        &= e^{-iH\Delta t} \ket{\psi(t')} \displaybreak[0] \\
        &= 0.706 e^{-0.331 i} e^{-i E_1 \Delta t} \psi_1 + 0.706 e^{-0.331 i} e^{-i E_2 \Delta t} \psi_2 + 0.060 e^{i E_3 \Delta t} \psi_3 \displaybreak[0] \\
        &= 0.706 e^{-0.331 i} e^{-i (-1 - \sqrt{2/N}) \Delta t} \frac{1}{\sqrt{2}} \left( \ket{a} + \ket{b} \right) + 0.706 e^{0.331 i} e^{-i (-1 + \sqrt{2/N}) \Delta t} \frac{1}{\sqrt{2}} \left( -\ket{a} + \ket{b} \right) + 0.060 e^{-i(N/2)t} \ket{e} \displaybreak[0] \\
        &= e^{it} \Bigg[ \frac{0.706}{\sqrt{2}} \left( e^{i (\sqrt{2/N} \Delta t - 0.331)} - e^{-i(\sqrt{2/N} \Delta t - 0.331)} \right) \ket{a} + \frac{0.706}{\sqrt{2}} \left( e^{i (\sqrt{2/N} \Delta t - 0.331)} + e^{-i (\sqrt{2/N} \Delta t - 0.331)} \right) \ket{b} \\
            &\quad\quad\quad + 0.060 \ket{e} \Bigg] \displaybreak[0] \\
        &= e^{it} \left[ 0.706 \sqrt{2} i \sin \left( \sqrt{\frac{2}{N}} \Delta t - 0.331 \right) \ket{a} + 0.706 \sqrt{2} \cos \left( \sqrt{\frac{2}{N}} \Delta t - 0.331 \right) \ket{b} + 0.060 \ket{e} \right].
\end{align*}
\end{widetext}
Note the global phase of $e^{it}$ can be dropped. Taking the norm-square of each amplitude, the probability of finding the particle in each type of vertex is
\begin{align*}
    p_a(\Delta t) &= 0.996 \sin^2 \left( \sqrt{\frac{2}{N}} \Delta t - 0.331 \right), \displaybreak[0] \\
    p_b(\Delta t) &= 0.996 \cos^2 \left( \sqrt{\frac{2}{N}} \Delta t - 0.331 \right), \displaybreak[0] \\
    p_c(\Delta t) &= 0, \displaybreak[0] \\
    p_d(\Delta t) &= 0, \displaybreak[0] \\
    p_e(\Delta t) &= 0.004.
\end{align*}
The success probability $p_a{(\Delta t)}$ reaches its maximum value of
\begin{equation}
    \label{eq:twostage-prob}
    p_* = 0.996
\end{equation}
when the argument to the sine function is $\pi/2$, i.e.,
\[ \sqrt{\frac{2}{N}} \Delta t - 0.331 = \frac{\pi}{2}. \]
Solving for $\Delta t$, the second stage of the algorithm must run for time
\begin{equation}
    \label{eq:stage2-runtime}
    \Delta t = \left( \frac{\pi}{2} + 0.331 \right) \sqrt{\frac{N}{2}} \approx 1.345 \sqrt{N}.
\end{equation}

To summarize, the first stage of the algorithm takes time $t' \approx 3.265\sqrt{N}$ \eqref{eq:stage1-runtime} for a probability of $0.996$ \eqref{eq:stage1-prob} to collect at the marked clique. Then, the second stage of the algorithm takes time $\Delta t \approx 1.345 \sqrt{N}$ for the probability of $0.996$ at the marked clique to collect at the marked vertex. Adding the runtimes of both stages, the total evolution time for the two-stage algorithm is
\begin{align}
    t_*
        &= t' + \Delta t \nonumber \displaybreak[0] \\
        &= 3.265 \sqrt{N} + 1.345 \sqrt{N} \nonumber \displaybreak[0] \\
        &= 4.610 \sqrt{N}. \label{eq:twostage-runtime}
\end{align}
The peak success probability of $0.996$ is independent of $N$, as verified by \fref{fig:Adjacency-twostage-Nvaries}, where we have plotted the two-stage search algorithm for various values of $N$. They all reach the same height of $0.996$. This figure also confirms the runtime expression \eqref{eq:twostage-runtime}, since the solid black curve peaks at time $4.610 \sqrt{1024} \approx 147.5$ (as also seen in \fref{fig:Adjacency-twostage-N1024}), the dashed red curve peaks at time $4.610 \sqrt{2048} \approx 208.6$, and the dotted green curve peaks at time $4.610 \sqrt{4096} \approx 295.0$.

%-------------------------------------------------------------------------------
% Section
%-------------------------------------------------------------------------------

\section{\label{sec:conclusion}Conclusion}

Previous research exploring search on the weighted truncated simplex lattice and linked complete graphs have shown that the weights between the cliques could be tuned to improve the runtime and success probability of quantum walk search algorithms by allowing probability to be freed from cliques. In these, the Laplacian and adjacency quantum walks searched in the same manner. In this paper, we explored the most restrictive means to join two cliques by investigating the weighted barbell graph of $N$ vertices, which has a single bridge of weight $w$. The Laplacian and adjacency quantum walks generally search this graph differently. For the Laplacian quantum walk, the weighted edge makes no difference; the evolution is the same regardless of the weight, where the probability is confined in each clique, and so the success probability reaches a maximum value of $1/2$ at time $\pi\sqrt{N}/2\sqrt{2}$. That is, the single bridge is too restrictive to affect the Laplacian quantum walk. For the adjacency quantum walk, most weights also make no difference, with the system evolving just like the Laplacian quantum walk. When $w = N/2$, however, the adjacency quantum walk does free some of the probability that was trapped in the unmarked clique, allowing the success probability to reach a value of 0.820 at time $2.518 \sqrt{N}$, where the coefficient is the solution to a transcendental equation. Evolving beyond this, to a time of $3.265 \sqrt{N}$, decreases the success probability, but causes 99.6\% of the total probability to be in the marked clique. Then, the weight can be changed, for example to $w = 1$, and the system will evolve to focus the success probability to the marked vertex, which takes an additional time of $1.345 \sqrt{N}$. Altogether, this two-stage algorithm takes $4.610 \sqrt{N}$ time to reach a success probability of $0.996$. Although this two-stage algorithm takes longer than search on the unweighted graph with repetitions, it could be useful in a scenario where it is expensive to repeat an algorithm, beyond the cost of querying the oracle.

%-------------------------------------------------------------------------------
% Acknowledgments
%-------------------------------------------------------------------------------

\begin{acknowledgments}
	This material is based upon work supported in part by the National Science Foundation EPSCoR Cooperative Agreement OIA-2044049, Nebraska’s EQUATE collaboration. Any opinions, findings, and conclusions or recommendations expressed in this material are those of the author(s) and do not necessarily reflect the views of the National Science Foundation.
\end{acknowledgments}

%-------------------------------------------------------------------------------
% References.
%-------------------------------------------------------------------------------

\bibliography{refs}

\end{document}